# Investigation of microwave dielectric relaxation process in the antiferroelectric phase of $NaNbO_3$ ceramics

M. H. Lente,[1] J. de Los S. Guerra and J. A. Eiras

Universidade Federal de São Carlos - Departamento de Física

Grupo de Cerâmicas Ferroelétricas - CEP 13565-670 - São Carlos - SP - Brazil

S. Lanfredi

Faculdade de Ciências e Tecnologia - FCT - Departamento de Física, Química e Biologia

Universidade Estadual Paulista - UNESP - CEP 19060-900 - Presidente Prudente - SP - Brazil

## Abstract

This letter reports microwave dielectric measurements performed in the antiferroelectric phase of $NaNbO_3$ ceramics from 100 to 450 K. Remarkable dielectric relaxations were found within the antiferroelectric phase and in the vicinity of ferroelectric-antiferroelectric phase transition. Such dielectric relaxation process was associated with relaxations of polar nanoregions with strong relaxor-like characteristic. In addition, the microwave dielectric measurements also revealed an unexpected and unusual anomaly in the relaxation strength, which was related to a disruption of the antiferroelectric order induced by a possible AFE-AFE phase transition.

[1] E-mail address: mlente@df.ufscar.br

# 1. Introduction

Ferroelectrics emerge from the dielectric material class as the highest dielectric constant materials, even when compared to other polar systems [1]. In order to investigate the electrical permittivity response of ferroelectric materials, dielectric measurements covering a wide frequency range (dielectric dispersion) are often required [2]. Dielectric dispersion measurements in both "normal" ferroelectrics and relaxor ferroelectrics (relaxors) have revealed remarkable dielectric relaxations in the microwave region [3, 4, 5, 6]. Thus, several efforts have been made in order to identify the mechanisms responsible for such relaxations. However, the exact origin is still debated and continues to be fascinating puzzle. For instance, it is proposed that the main origin of microwave relaxation process in "normal" ferroelectrics is related to the relaxation of 90° domain walls [7, 8]. On the other hand for relaxors, the relaxation mechanism has been attributed to the flipping or fluctuations of the boundaries of polar clusters [9, 10].

Recently, antiferroelectric materials such as La-modified lead zirconate titanate [11] and sodium niobate [12, 13] ($NaNbO_3$) with relaxor-like phase transition have emerged as promising and intriguing new systems. Particularly, $NaNbO_3$ has received special attention due to its high complexity. Indeed, $NaNbO_3$ has presented great interest due to the ability to present successive phase transitions from non-polar (PE) to antiferroelectric (AFE) and finally reaching a ferroelectric phase (FE) [14]. It is reported that the coupling between AFE and FE order parameters sometimes associated with intrinsic defects is responsible for remarkable dielectric features in $NaNbO_3$ [15, 16]. However, detailed investigation of the dielectric properties of $NaNbO_3$ in the microwave region in the antiferroelectric and ferroelectric phases is still missing. In the best knowledge of the authors, there are no works in the current literature reporting the microwave dielectric properties in antiferroelectric systems.

The aim of this letter is to characterize the electrical permittivity of $NaNbO_3$ ceramics through a dielectric dispersion in a wide range of frequency and temperature in order to investigate the origin of the dielectric relaxations in this antiferroelectric system. Thus, it is believed that the present work can contribute to a better understanding of the physical phenomena involved in the microwave relaxation process in ferroelectric materials.

## 2. Experimental section

$NaNbO_3$ powder was prepared by a new chemical route using the thermal decomposition of a precursor salt synthesized by solution evaporation method. This procedure was described in details in previous work [12, 17] High frequency measurements were performed using a Network Analyzer HP-8719C in a frequency and temperature range from 50 MHz to 2 GHz and from 100 K to 450 K, respectively. Low frequency measurements were performed using an Impedance Analyzer HP 4194A in a frequency and temperature range from 1 kHz to 1 MHz and from 15 K to 870 K, respectively.

## 3. Results and discussion

Figures 1 show the real (ε') and the imaginary (ε'') parts of the electrical permittivity as function of the temperature and frequency obtained during the cooling cycle at 2 K/min. The results in figure 1 reveal two clear very broad dielectric anomalies (peaks) occurring around 100 K (I) and 300 K (II) at 1 kHz with a slight thermal hysteresis of 2 K (for the sake of simplicity not shown here). The origin of the first anomaly (I) may be associated with antiferroelectric-ferroelectric phase transition, as reported previously [12, 14]. However, the origin of the second one (II) at 300 K is not still clear but it has been associated with a possible antiferroelectric-antiferroelectric phase transition [12, 13]. Moreover, these results remarkably reveal that both anomalies present a relaxor-type behavior.

The frequency dependence of the electrical permittivity was characterized from 100 K to 450 K, thus covering the second anomaly (II) and the vicinity of the first one (I). The results for ε´ and ε´´ measured at room temperature are shown in figure 2 (a). The data reveal two dielectric relaxations. The first one occurs at lower frequencies being very weak and broad, covering almost four decades and centered around 740 kHz (see inset for better visualization). However, the results remarkably revealed that this relaxation occurs only at 300 K, being suppressed just below and above it. On the other hand, the second dielectric relaxation is strong and sharp, being noticed in the microwave region, which relaxation frequency is 944 MHz. It is also verified that from 100 K to 450 K the dielectric relaxation in the microwave region is always observed. Figure 2 (b) shows some representative curves of ε' and ε" measured at different temperatures in the GHz range.

The temperature dependence of the relaxation frequency ($f_R$) and the relaxation strength (Δε) obtained from the characterizations in the microwave range are presented in figure 2 (c). The results reveal that $f_R$ decreases with increasing the temperature and a minimum is reached around 350 K. Then, $f_R$ increases subtlety with the increase of the temperature. A change in the slope of $f_R$ curve is noticed at 300 K. On the other hand, Δε increases with the increase of the temperature up to 275 K but it drops abruptly reaching a minimum at 300 K. After that, Δε increases again and reaches a maximum at 350 K, which coincides with the minimum of $f_R$, then decreasing slowly with the increase of the temperature. It is remarkable that the sudden drop of Δε and its minimum at 300 K coincide with both the change of the slope of $f_R$ and that weak relaxation observed at lower frequencies (figure 2(a)). These results were reproducible and identical in the cooling and heating cycles. Microwave dielectric relaxations in antiferroelectric systems as well as this unusual anomaly observed in $\Delta\varepsilon$ at 300 K have never been reported before. Thus, these novel features will be the focus of this work.

As commented in the introduction, the physical origin of dielectric relaxations in the microwave region for "normal" ferroelectrics and for relaxors has been analyzed distinctly and attributed respectively to domain walls [7] and to nanometer polar regions [9]. However, in our case two fundamental points must be emphasized. First, $NaNbO_3$ is an antiferroelectric system, thus presenting well-defined antiferroelectric domain structures [14]. And second, the dielectric measurements also reveal a relaxor-like characteristic (figure 1) that suggests strongly the presence of nanopolar regions with short-range order [18, 19]. Therefore, there is a great complexity to determine the mechanism responsible for the dielectric relaxation in the GHz range for $NaNbO_3$ since in principle, both antiferroelectric domain structures and polar nanoregions could be responsible for such relaxation process. Thus, the fundamental task is to identify which of them is the dominant mechanism.

It is reported in the literature that for "normal" ferroelectrics $f_R$ presents an abrupt decrease when the temperature increases, thus reaching a well defined minimum in a temperature that coincides with the Curie temperature ($T_C$) [20, 21]. On the other hand, it has been reported that for relaxors $f_R$ decreases with the increase of the temperature reaching a minimum and tending to level out in a temperature that coincides with the temperature of maximum permittivity ($T_m$) [20, 22] similarly as observed in our results in figure 2 (c). Therefore, the fact that the relaxation frequency of $NaNbO_3$ becomes reasonably independent of the temperature at temperatures higher than 350 K allows us to infer that the mechanism responsible for the dielectric relaxation in the microwave range may be associated with relaxation of polar nanoregions rather than antiferroelectric domain or domain wall structures, thus prevailing the relaxor characteristic. This hypothesis corroborates the dielectric results observed in figure 1.

Nevertheless, the physical origin of the anomaly of $\Delta\varepsilon$ in the vicinity of 300 K is unclear. There are no theories in the literature to explain such behavior. Curiously it occurs near the temperature interval where is hypothesized to occur an AFE-AFE phase transition. Therefore, it is inferred that around 300 K occurs a kind of disruption in the antiferroelectric range order due to the supposed AFE-AFE phase transition that induces the anomaly observed in $\Delta\varepsilon$, which would be also responsible for that relaxation in the radio frequency. However, further detailed investigation is required to confirm this supposition.

## 4. Conclusion

In summary, the electrical permittivity of $NaNbO_3$ ceramics was investigated in a temperature interval covering the antiferroelectric phase and the vicinity of the ferroelectric-antiferroelectric phase transition. The results showed remarkable dielectric relaxations in the microwave region, which mechanism was associated with the dynamics of nanopolar regions with relaxor feature. In addition, an unexpected anomaly in $\Delta\varepsilon$ was observed around 300 K and related to a disruption in the range order due to an AFE-AFE phase transition.

Acknowledgements

The authors thank FAPESP, CAPES and CNPq for financial support.

# Figure Captions

Figure 1: Temperature and frequency dependence of the real (ε') and the imaginary (ε'') parts of the electrical permittivity.

Figure 2: (a) frequency dispersion of the electrical permittivity performed at room temperature in all frequency range investigated; (b) ε' and ε'' characterized in the microwave region at different temperatures; (c) temperature dependence of $f_R$ and Δε.

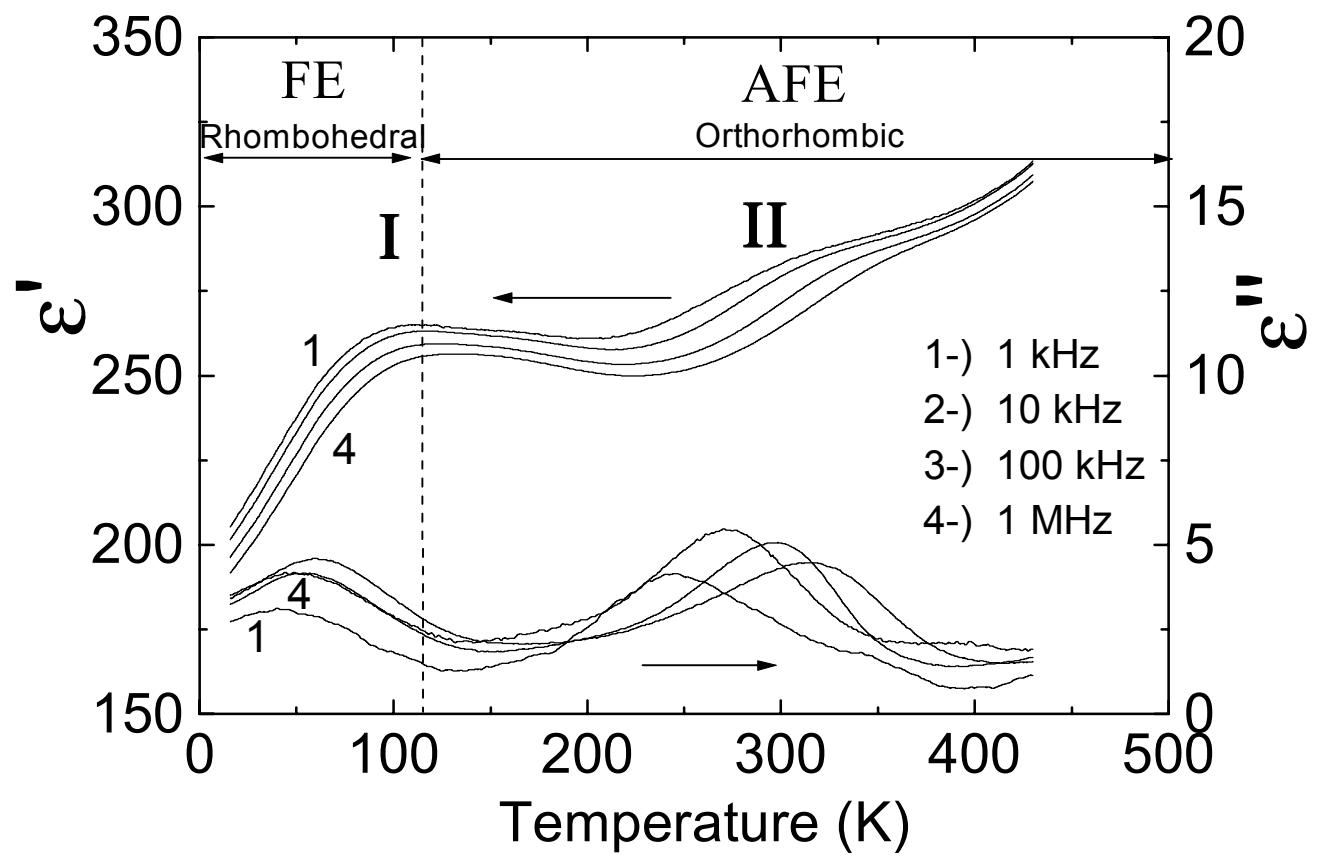

Figure 1: M. H. Lente et. al – Appl. Phys. Lett.

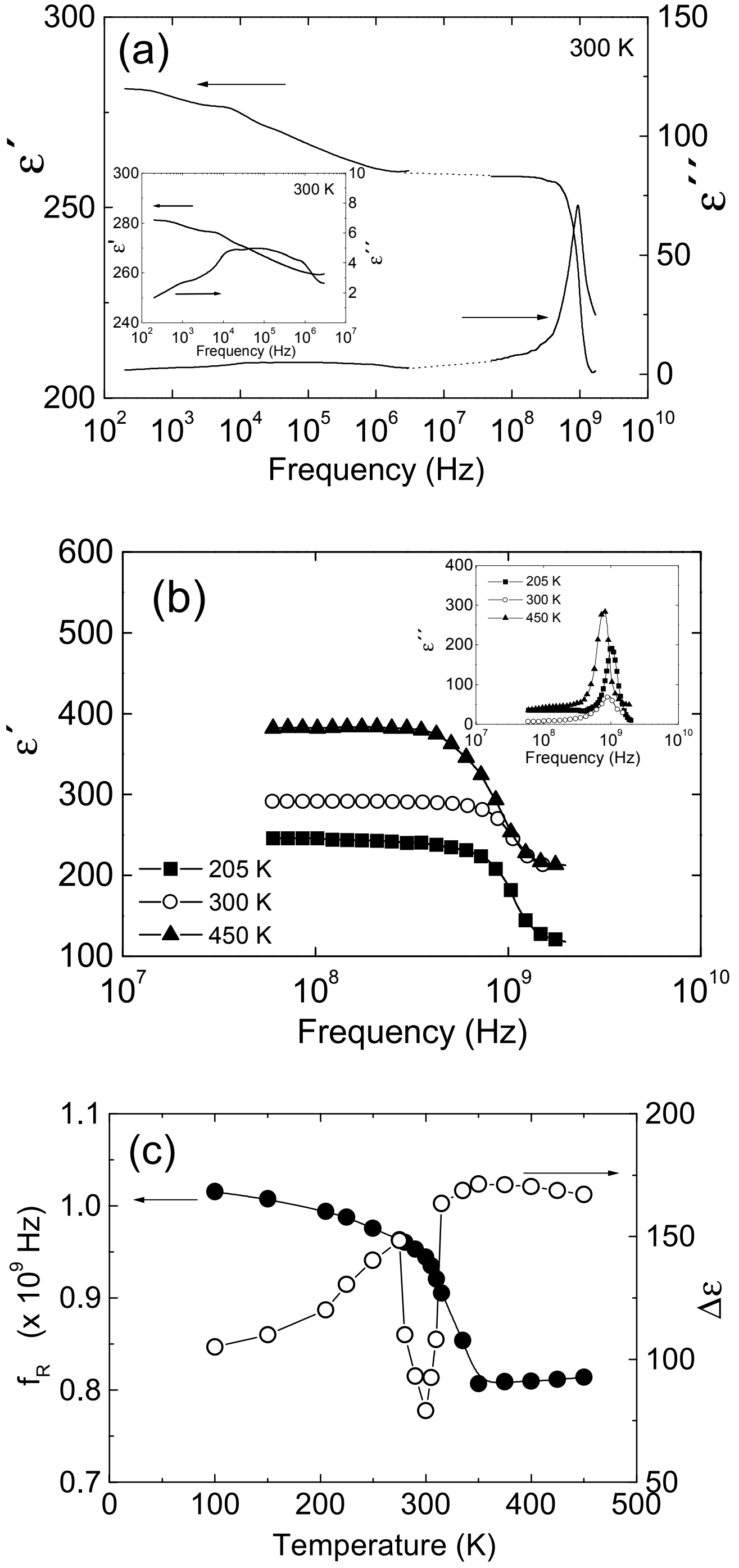

Figure 2: M. H. Lente et. al – Appl. Phys. Lett.